\documentclass[technote,10pt]{IEEEtran}
\usepackage{caption}
\usepackage[nocompress]{cite}
\usepackage{amsmath}
\usepackage{graphicx}
\usepackage{picinpar}
\usepackage{authblk}
\usepackage{amsfonts,amssymb}
\usepackage{mdwmath}
\usepackage{mdwtab}
\usepackage{eqparbox}
\usepackage{mathtools}
\usepackage{subcaption}
\usepackage{ragged2e}
\usepackage{extarrows}
\usepackage{booktabs}
\usepackage{multirow}
\usepackage{tabularx}
\usepackage{indentfirst}
\usepackage{makecell}
\usepackage[justification=centering]{caption}
\usepackage{siunitx}    
\usepackage{algorithmic}
\usepackage{xcolor}
\usepackage{bm}
\usepackage[margin=1.6cm]{geometry}
\usepackage{setspace}  
\usepackage{enumitem}   
\usepackage{longtable}
\usepackage[mathscr]{euscript}
\usepackage{array}
\usepackage[colorlinks=true, allcolors=black]{hyperref}
\usepackage{stfloats}
\usepackage{etoolbox} 
\usepackage{titlesec}
\titlespacing*{\section}{0pt}{0pt}{0pt} 
\titlespacing*{\subsection}{0pt}{0pt}{0pt}
\titlespacing*{\subsubsection}{0pt}{0pt}{0pt}
\usepackage[switch]{lineno} 
\usepackage[linesnumbered,ruled,vlined]{algorithm2e}
\makeatletter

\def\BibColor@get#1{%
	\@ifundefined{BibColor@#1}{}{%
		\csname BibColor@#1\endcsname}%
}

\AtBeginDocument{%
	\let\BC@orig@lbibitem\@lbibitem
	\let\BC@orig@bibitem \@bibitem
	
	\def\@bibitem#1{%
		\edef\BC@col{\BibColor@get{#1}}%
		\ifx\BC@col\@empty
		\color{black}
		\else
		\color{\BC@col}
		\fi
		\BC@orig@bibitem{#1}%
	}%
	
	\def\@lbibitem[#1]#2{%
		\edef\BC@col{\BibColor@get{#2}}%
		\ifx\BC@col\@empty
		\color{black}%
		\else
		\color{\BC@col}%
		\fi
		\BC@orig@lbibitem[#1]{#2}%
	}%
}
\makeatother

\begin{document}
\setlength{\textfloatsep}{1pt}
\setlength{\abovecaptionskip}{2pt}
\setlength{\floatsep}{-1pt}
\newtheorem{lemma}{Lemma}
\newtheorem{corol}{Corollary}
\newtheorem{theorem}{Theorem}
\newtheorem{proposition}{Proposition}
\newtheorem{definition}{Definition}
\newcommand{\e}{\begin{equation}}
\newcommand{\ee}{\end{equation}}
\newcommand{\eqn}{\begin{eqnarray}}
\newcommand{\eeqn}{\end{eqnarray}}
\newcommand{\figsub}[2]{Fig.~\ref{#1}(\subref{#2})}
\renewcommand{\algorithmicrequire}{\textbf{Input:}} 
\renewcommand{\algorithmicensure}{\textbf{Output:}} 
\renewcommand{\raggedright}{\leftskip=0pt \rightskip=0pt plus 0cm}

\title{Multi-Domain Iterative Detection for Massive Connectivity in LEO Satellite Networks}

\author{Xinhua Liu, Yueqing Wang, Keke Ying, Peihu Duan, Dapeng Li, Ziwei Wan, Zhongliang Zhao, Chabalala S. Chabalala, Andrey Ivanov, and Zhen Gao
\vspace*{-3.0mm}

\thanks{The work was supported in part by Beijing Natural Science Foundation under Grants L242011, in part by the Natural Science Foundation of China (NSFC) under Grant 62471036 and Grant U2233216, in part by Shandong Province Natural Science Foundation under Grant ZR2025QA30, in part by Fundamental Research Funds for the Central Universities.  \emph{(Corresponding authors: Zhen Gao and Dapeng Li.)}}
\thanks{Xinhua Liu, Yueqing Wang, Keke Ying, and Ziwei Wan are with the School of Information and Electronics, Beijing Institute of Technology, Beijing 100081, China (e-mails: xinhualiu@bit.edu.cn; 3120232044@bit.edu.cn; ykk@bit.edu.cn; ziweiwan@bit.edu.cn).}
\thanks{Peihu Duan is with the School of AI, Beijing Institute of Technology, Beijing 100081, China (e-mail: duanpeihu@bit.edu.cn).}
\thanks{Dapeng Li is with the School of Optics and Photonics, Beijing Institute of Technology, Beijing 100081, China (e-mail: dapangli@bit.edu.cn).}
\thanks{Zhongliang Zhao is with the School of Electronic Information Engineering, Beihang University, Beijing 100191, China (e-mail: zhaozl@buaa.edu.cn).}
\thanks{Zhen Gao is with the State Key Laboratory of CNS/ATM, Beijing 100081, China, also with the MIIT Key Laboratory of Complex-Field Intelligent Sensing, Beijing 100081, China, also with BIT, Zhuhai 519088, China, also with the Advanced Technology Research Institute, BIT, Jinan 250307, China, also with the Yangtze Delta Region Academy, BIT, Jiaxing 314019, China, and also with the Shaanxi Key Laboratory of Information Communication Network and Security, Xi'an University of Posts $\&$ Telecommunications, Xi'an, Shaanxi 710121, China (e-mail: gaozhen16@bit.edu.cn).}
\thanks{Chabalala S. Chabalala is with the School of Electrical and Information Engineering, University of the Witwatersrand Johannesburg, South Africa (e-mail: chabalala.chabalala@wits.ac.za).}
\thanks{A. Ivanov is with the Internet of Things and Wireless Technologies Center, Skolkovo Institute of Science and Technology (Skoltech), Moscow, Russia (e-mail: AN.Ivanov@skoltech.ru).}
}
\IEEEpubid{}

\maketitle

\begin{abstract}
Grant-Free (GF) random access is promising for low Earth orbit satellite Internet due to its reduced access latency. However, existing schemes suffer from poor performance in massive connectivity scenarios. To address this challenge, we firstly propose an iterative residual feedback multi-measurement vector approximate message passing algorithm. This algorithm leverages multi-domain synergistic sparsity in the spatial-frequency and angular-delay domains to alternately perform active user terminal detection (AUD) and channel estimation (CE). Additionally, a residual feedback mechanism is incorporated to suppress error accumulation, thereby enhancing AUD performance. Furthermore, conventional data detection (DD) methods significantly degrade when active user terminals are spatially close or outnumber the satellite's receive antennas, making the demodulation problem rank-deficient or underdetermined. To mitigate this, we design a data modulation scheme via joint spatial-frequency multi-domain spreading, which utilizes observations from both spatial and frequency domains to facilitate multi-domain DD. Simulation results demonstrate that the proposed scheme significantly outperforms existing GF methods in terms of AUD accuracy, CE precision, and bit error rate, especially under conditions of low effective pilot length and practical signal-to-noise ratios.

\end{abstract}
\IEEEpeerreviewmaketitle

\section{Introduction}
Low Earth orbit (LEO) satellites are pivotal for realizing the global coverage vision of sixth-generation (6G) networks~\cite{LEO_seamless}. With the advent of the 6G era, massive machine-type communications-satellite (mMTC-s) has become a key requirement to support ubiquitous connectivity~\cite{ITU}. While terrestrial networks serve urban areas, a vast number of Internet-of-Things (IoT) devices distributed in remote regions require satellite access to bridge the coverage gap. Consequently, LEO satellite networks must support massive connectivity for these terminals. In such scenarios, traditional grant-based handshake protocols are unsuitable due to their excessive signaling overhead and high latency for bursty traffic~\cite{2021_Kodheli_survey}. Therefore, grant-free random access (GF-RA), which bypasses the complex handshake procedure, is adopted to effectively enhance access capacity and reduce latency~\cite{2020_Shahab_GFRA}, making it promising for satellite IoT~\cite{2023_Zhou}. Furthermore, to address the massive connectivity challenge, GF non-orthogonal multiple access (NOMA) has emerged as a promising solution. In particular, code-domain NOMA schemes have been extensively investigated to enhance overloading capabilities by assigning unique non-orthogonal spreading sequences to user terminals (UTs)~\cite{CD-NOMA}.

To efficiently identify these active UTs and recover their signals from the superimposed non-orthogonal codes, the framework of joint user activity detection and channel estimation (JADCE) has been adopted in works such as~\cite{2023_Zuo_LEO} and~\cite{2023_Ying_LEO} to identify active UTs and estimate channel state information simultaneously. In terms of waveforms, schemes based on single-carrier~\cite{2020_Zhang_BR-MP-EM}, orthogonal frequency division multiplexing (OFDM)~\cite{2023_Zuo_LEO}, and orthogonal time frequency space (OTFS)~\cite{2023_zxy_LEO} have been proposed to handle Doppler shifts and delay spreads. However, these methods often employ greedy algorithms, such as simultaneous orthogonal matching pursuit (SOMP)~\cite{2023_zxy_LEO}, which fail to exploit channel priors or effectively handle high inter-user correlation in overloaded scenarios. While multi-satellite cooperation~\cite{2023_Ying_LEO} exploits spatial diversity to improve detection performance, it entails high inter-satellite link overhead and synchronization challenges, and often neglects the synergistic gain of multi-domain sparsity.

Additionally, since the implementation of massive multiple-input multiple-output (mMIMO) on satellites is limited by huge deployment costs and power consumption, another fundamental bottleneck remains in LEO-based massive access: the number of active terminals often far exceeds the number of onboard antennas, resulting in a severely underdetermined system that traditional zero-forcing multi-user data detection (DD) cannot resolve. To address such overloaded detection problems, various receiver processing schemes have been proposed. Methods such as linear and interference-cancellation-based detection\cite{2014_Qian_linear}, nonlinear detection\cite{2017_unlinear_detection}, message passing\cite{2019_GMP}, and sparse recovery\cite{2016_Gao_CS} predominantly treat the overloaded condition as a receiver-end inference problem under a predetermined observation model. They aim to boost performance primarily by enhancing the algorithms' recovery capabilities for underdetermined systems, thereby remaining confined to receiver-side enhancements within a fixed observation dimension. However, constrained by limited onboard antennas and stringent size, weight, and power limitations, these approaches struggle to fully adapt to LEO scenarios. In particular, under conditions of highly correlated angles of arrival (AoA), detection schemes relying solely on spatial discrimination often suffer from significant performance degradation due to severely ill-conditioned channels. Consequently, it is imperative to construct a joint design framework that integrates observation dimension expansion with multi-domain structure utilization, fundamentally alleviating the underdetermined bottleneck inherent in spatial-domain-only  detection.

To this end, we propose to further explore the sparsity of angular- and delay-domain access channels for enhanced active UT detection (AUD) performance, and conceive a data modulation via spatial-frequency multi-domain spreading to prevent the demodulation problem from becoming rank-deficient. Our contributions are as follows:
\vspace{-3pt}
\begin{itemize}
    \item Multi-Domain Synergistic Strategy: We develop a multi-domain synergistic framework that alternately executes AUD in the spatial-frequency domain (SF-domain) and channel estimation (CE) in the angular-delay domain (AD-domain). This approach fully exploits the structured sparsity of the SF-domain and the enhanced cluster sparsity of the AD-domain.
    \item Spreading-based observation dimension expansion for data modulation: To address the rank-deficient demodulation problem inherent in the underdetermined massive connectivity system caused by severe overload and limited onboard antennas in LEO systems, we design a joint spatial-frequency spreading modulation scheme that expands the effective observation dimension and improves detection robustness under highly correlated AoAs.
    \item Residual Feedback Algorithm: We propose an iterative residual feedback multi-measurement vector approximate message passing (IRF-MAMP) algorithm. It incorporates a residual feedback mechanism to cancel interference from UTs with high activity probabilities for error suppression, followed by linear minimum mean square error (LMMSE)-based DD.
\end{itemize}

Simulation results confirm the effectiveness and superiority of the proposed approach in practical LEO satellite massive access scenarios.

\textit{Notations:} 
Tensors, matrices, and vectors are denoted by $\boldsymbol{\mathscr{H}}, \mathbf{H}, \mathbf{h}$, respectively. $\boldsymbol{\mathscr{H}}_{:,:,n}$, $\boldsymbol{\mathscr{H}}_{:,g,n}$, and $\boldsymbol{\mathscr{H}}_{k,g,n}$ denote a slice, a fiber, and the $(k,g,n)$-th element of a tensor, respectively. The operators $\otimes$, $\circ$, and $\times_n$ denote the Kronecker, Hadamard, and $n$-mode products, respectively. The cardinality of a set $\mathcal{K}$ is denoted by $|\mathcal{K}|_c$.

\section{System Model}
\label{system model}
We consider an uplink RA scenario in which the LEO satellite provides OFDM-based connectivity services to $K$ UTs that can access the satellite via very small aperture terminals. The satellite is equipped with a uniform planar array (UPA) of $N_r\!=\!N_r^x\!\times\! N_r^y$ antenna elements, where $N_r^x$ and $N_r^y$ denote the number of antennas along the $x$-axis and $y$-axis, respectively. Similarly, each UT is equipped with a UPA composed of $N_t\!=\!N_t^x\!\times\! N_t^y$ antenna elements. 

Due to the sporadic traffic nature of IoT applications, only $K_a (\ll \!K$) UTs are active, characterized by a binary indicator $\alpha_k\!\in\!\{\!0,\!1\!\}$. The set of all UTs is denoted by $\mathcal{K}$, and the active UT set (AUS) is defined as $\mathcal{K}_a\!=\!\{k|\alpha_k\!=\!1,\!1\!\leq\! k\!\leq\! K\}$, with the number of active UTs denoted by $|\mathcal{K}_a|_c$. The line-of-sight (LoS)-dominant channel exhibits minimal angular spread at the receiver due to limited local scatterers and the large satellite-UT distance. Therefore, we assume that all multipath components share a common angle of arrival (AoA) while exhibiting different propagation delays. Additionally, in the considered receiver architecture, Doppler compensation is treated as a front-end synchronization function. In particular, the dominant Doppler component is assumed to be pre-compensated using ephemeris information, and the remaining frequency offset is further reduced through receiver synchronization procedures~\cite{2023_Zuo_LEO, Doppler2, 2023_Ying_LEO}. Accordingly, the following model focuses on the post-synchronization AUD, CE, and DD procedures. For modeling simplicity, the residual Doppler impairment is absorbed into the equivalent receiver noise. Even after Doppler compensation, the LEO uplink remains challenging due to severe overload, limited onboard antennas, and highly correlated AoA.


The signal transmitted by UTs propagates through the Rician channel characterized by $L_p$ paths and a maximum number of $Q$ delay taps. Considering the uplink transmission with a total of $M$ subcarriers, the SF-domain channel matrix $\check{\mathbf{H}}\in\mathbb{C}^{N_r\times N_t}$ between the $k$-th UT and the LEO satellite at the $m$-th subcarrier can be modeled as
\begin{equation} \label{spacefrequency channel}
	\begin{aligned}
		\check{\mathbf{H}}_{k,m} \!&=\!  \alpha_k\!\sum\nolimits_{l_p\!=0}^{L_p\!-1} \tilde{\beta}_{k,l_p,m} \mathbf{a}_r\left( \theta_{k}, \phi_{k} \right) \mathbf{a}_t^{\mathsf{H}} \left( \theta_{k,l_p}\!, \phi_{k,l_p} \right),
	\end{aligned}
\end{equation}
where $l_p\!=\!0,1,\cdots, L_p\!-\!1$, $\tilde{\beta}_{k,l_p,m}=\beta_{k,l_p}\sum^{Q-1}_{q=0}\varrho(qT_s-\tau_{k,l_p})e^{j2\pi q(m-1)/M}$ with $q=0,1,\cdots, Q-1$. $\beta_{k,l_p}$,  $\varrho(\tau)$, and $\tau_{k,l_p}$ denote the complex path gain, a pulse shaping filter with $T_s$-spaced signaling, and the channel delay, respectively. Specifically, the path gains are defined as $\beta_{k,0}=\sqrt{\gamma/(\gamma+1)}$ for the LoS path ($l_p=0$) and $\beta_{k,l_p}\sim\sqrt{1/[(\gamma+1)(L_p-1)]}\mathcal{CN}(\beta;0,1)$ for the non-LoS paths ($l_p\neq0$), where $\gamma$ is the power distribution factor and $L_p\geq 1$, $\mathcal{CN}(\beta;0,1)$ denotes the complex Gaussian distribution of a random vector $\beta$ with zero mean and identity covariance matrix. The satellite steering vector $\mathbf{a}_r(\theta_{k}, \phi_{k})\in\mathbb{C}^{N_r\times 1}$ and the $k$-th UT steering vector $\mathbf{a}_t(\theta_{k,l_p}, \phi_{k,l_p})\in\mathbb{C}^{N_t\times 1}$ are defined as     $\mathbf{a}_r \left( \theta_{k}, \phi_{k} \right) = \mathbf{v}_k^x \left( \theta_{k}, \phi_{k} \right) \otimes \mathbf{v}_k^y \left( \theta_{k}, \phi_{k} \right)$ and $\mathbf{a}_t \left( \theta_{k,l_p}, \phi_{k,l_p} \right) = \mathbf{v}_k^x \left( \theta_{k,l_p}, \phi_{k,l_p} \right) \otimes \mathbf{v}_k^y \left( \theta_{k,l_p}, \phi_{k,l_p} \right)$, respectively. The components $\mathbf{v}_k^{\star} \left( \theta_{k}, \phi_{k} \right)=[1\quad e^{-j\mu_{k}^{\star}}\cdots e^{-j(N_{\star}-1)\mu_{k}^{\star}}]^{\mathsf{T}}$ for $\star\!\in\!\{x,y\}$, $\mu^{x}_{k}=(2\pi d/\lambda){\rm cos}\theta_{k}{\rm sin}\phi_{k}$, and $\mu_{k}^{y}=(2\pi d/\lambda){\rm sin}\theta_{k}{\rm sin}\phi_{k}$, depend on the wavelength $\lambda$, antenna spacing $d=\lambda/2$, and the AoAs $\theta_k, \phi_{k}$. 

Leveraging predictable orbital trajectories of LEO satellites, analog beamforming at the UT is approximated using a matched filter $\boldsymbol{\omega}_{k}= \mathbf{a}_{t}(\theta_{k,0},\,\phi_{k,0})$. The effective channel vector at the $m$-th subcarrier is denoted as $\mathring{\boldsymbol{\mathscr{H}}}_{k,m,:}=	\check{\mathbf{H}}_{k,m}\boldsymbol{\omega}_{k}\in\mathbb{C}^{N_r \times  1}$. We denote the aggregate SF-domain channel tensor for all $K$ UTs as $\mathring{\boldsymbol{\mathscr{H}}}\in\mathbb{C}^{K\times M\times N_r}$.
\enlargethispage{\baselineskip}
\section{Proposed Transmit Signal Design and Uplink Transmission Problem Formulation}
In the context of the channel model established in Section~\ref{system model}, angular overlap in high-density scenarios increases channel correlation, while the high dimensionality of mMIMO channels introduces substantial computational complexity and latency. These characteristics exacerbate the rank-deficient demodulation issue encountered in existing two-stage transmission schemes, particularly when the number of active UTs exceeds the number of the satellite antennas.

To address these challenges, a resource allocation scheme based on SF-domain spreading is proposed. The entire set of subcarriers is partitioned into $G$ resource blocks, each containing $\Delta=M/G$ subcarriers. A subcarrier group (SG) is then formed by uniformly sampling the spectrum, selecting subcarriers at the same relative position (e.g., the $\xi$-th) within each group. Within each SG, the same constellation symbol is transmitted over all $G$ subcarriers. 

For the $n$-th receive antenna, the frequency-domain channel matrix for a selected SG, $\boldsymbol{\mathscr{H}}_{:,:,n}\in\mathbb{C}^{K\times G}$, is extracted from the full-band channel matrix $\mathring{\boldsymbol{\mathscr{H}}}_{:,:,n}\in\mathbb{C}^{K\!\times\! M}$ using a subcarrier selection matrix $\mathbf{V}\in\{0,1\}^{M\!\times\! G}$. This matrix is constructed as
\begin{align}
\boldsymbol{\mathscr{H}}_{:,:,n}\!=\!\mathring{\boldsymbol{\mathscr{H}}}_{:,:,n}\!\mathbf{V}\!=\! \left[\mathring{\boldsymbol{\mathscr{H}}}_{:,\xi,n},\!\mathring{\boldsymbol{\mathscr{H}}}_{:,\xi+\Delta,n}\!,\cdots\!,\!\mathring{\boldsymbol{\mathscr{H}}}_{:,\xi+(G\!-\!1)\Delta,n}\!\right]\!,
\end{align}
where the selection matrix $\mathbf{V}\!$ has entries $\mathbf{V}_{m,g}\! = \!1$ if $m\! = \xi + (g-1)\Delta$ and $0$ otherwise. Here, $\boldsymbol{\mathscr{H}}\in\mathbb{C}^{K\times G\times N_r}$ is the frequency-domain channel tensor for an SG.

Based on the proposed SF-domain spreading design, the uplink transmission is organized into a frame structure to facilitate practical random access. We incorporate our modulation scheme into a two-phase uplink transmission structure. Each frame consists of 
$T_p+T_d$ time slots: the first $T_p$ slots are used for pilot transmission and the remaining $T_d$ slots for data transmission. We denote the number of time slots under consideration by $T\in\{T_p,T_d\}$.

To enable user-specific signal separation, the $k$-th UT's signal is spread over $G$ subcarriers using a non-orthogonal spreading code $\mathbf{s}_{k}\in\mathbb{C}^{G\!\times\! 1}$. 
 We present the signal received by the LEO satellite as a tensor $\boldsymbol{\mathscr{Y}}^{b}\in\mathbb{C}^{T\times G\times N_r}$. Therefore, the signal received at the $n$-th antenna in the $t$-th time slot can be expressed as follows:
\begin{equation}
 \begin{aligned}
        \boldsymbol{\mathscr{Y}}^{b}_{t,:,n}&=\sum\nolimits^{K}_{k=1}(\boldsymbol{\mathscr{H}}_{k,:,n}\circ\mathbf{s}_k)x^{b}_{t,k}+\boldsymbol{\mathscr{N}}_{t,:,n},
 \end{aligned}
\end{equation}
where the superscript $b\!\in\!\{p, d\}$ indexes the pilot ($p$) and data ($d$) phases. The subchannel between the $n$-th satellite antenna and the $k$-th UT is $\boldsymbol{\mathscr{H}}_{k,:,n}\in\mathbb{C}^{G\times 1}$, $\boldsymbol{\mathscr{N}}_{t,:,n}\in\mathbb{C}^{G\times 1}$ denotes the additive white Gaussian noise (AWGN) at the $n$-th antenna and the $t$-th time slot, and $x^b_{t,k}$ denotes pilot or data symbols transmitted by the $k$-th UT at the $t$-th time slot. 

After derivation, we obtain
\begin{equation}
	\begin{aligned}
		\boldsymbol{\mathscr{Y}}^{b}_{t,:,n}
		=(\boldsymbol{\mathscr{H}}_{:,:,n}\circ\mathbf{S})^{\mathsf{T}}\mathbf{x}^{b}_{t}+\boldsymbol{\mathscr{N}}_{t,:,n}
		=\boldsymbol{\mathscr{E}}_{:,:,n}^{\mathsf{T}}\mathbf{x}^{b}_t+\boldsymbol{\mathscr{N}}_{t,:,n},
	\end{aligned}
\end{equation}
where $\mathbf{S}\!=\![\mathbf{s}_1, \mathbf{s}_2, \cdots, \mathbf{s}_K]^{\mathsf{T}}\!\in\!\mathbb{\!C\!}^{K\!\times\! G}$ and $\mathbf{x}_t^b\!=\![x^b_{t,1},x^b_{t,2},\cdots,x^b_{t,K}]^{\mathsf{T}}\!\in\!\mathbb{C}^{K\times 1}$. Define $\boldsymbol{\mathscr{E}}_{:,:,n}\!=\!\boldsymbol{\mathscr{H}}_{:,:,n}\circ\mathbf{S}\!\in\!\mathbb{C}^{K\times G}$ as the equivalent channel matrix; $\boldsymbol{\mathscr{E}}\in\mathbb{C}^{K\times G\times N_r}$ denotes the tensor whose $n$-th slice is $\boldsymbol{\mathscr{E}}_{:,:,n}$. We assume the transmitted signal power satisfies $\mathbb{E}[|x^b_{t,k}|^2]=1$.

At the LEO satellite, the signals received in $T$ successive time slots at the $n$-th antenna are organized as 
\begin{equation}\label{spa-fre-Y}
	\begin{aligned}        \boldsymbol{\mathscr{Y}}^b_{:,:,n}=\mathbf{X}^b\boldsymbol{\mathscr{E}}_{:,:,n}+\boldsymbol{\mathscr{N}}_{:,:,n},
	\end{aligned}
\end{equation}
where $\mathbf{X}^b=[\mathbf{x}_1^b,\mathbf{x}_2^b,\cdots,\mathbf{x}_{T}^b]^{\mathsf{T}}\in\mathbb{C}^{{T}\!\times\! K}$, $\boldsymbol{\mathscr{Y}}^b_{:,:,n}=[\boldsymbol{\mathscr{Y}}^{b}_{1,:,n},\boldsymbol{\mathscr{Y}}^{b}_{2,:,n},\cdots,\boldsymbol{\mathscr{Y}}^{b}_{T,:,n}]^{\mathsf{T}}$, and $\boldsymbol{\mathscr{N}}_{:,:,n}=[\boldsymbol{\mathscr{N}}_{1,:,n},\boldsymbol{\mathscr{N}}_{2,:,n},\cdots,\boldsymbol{\mathscr{N}}_{T,:,n}]^{\mathsf{T}}$. Due to the unique uplink propagation environment, the channel matrix $\boldsymbol{\mathscr{E}}_{:,:,n}$ exhibits a multi-level sparse structure. First, the traffic of UTs is inherently sparse, as only a small fraction of UTs are active. This leads the channel matrix $\boldsymbol{\mathscr{E}}_{:,:,n}$ on the $g$-th subcarrier to be sparse, and all subcarriers exhibit the same sparsity, i.e.,
\begin{equation}\label{fre sparity}
    \begin{aligned}        
     |{\rm supp}\{\boldsymbol{\mathscr{E}}_{:,1,n}\}|_c=\cdots=|{\rm supp}\{\boldsymbol{\mathscr{E}}_{:,G,n}\}|_c=K_a\ll K.
    \end{aligned}
\end{equation}
where ${\rm supp\{\cdot\}}$ is the support set of a matrix. Furthermore, since $\alpha_k$ remains constant for all subchannels, it induces a common spatial-domain sparsity pattern across $\{\boldsymbol{\mathscr{E}}_{:,:,n}\}^{N_r}_{n=1}$, as follows
\begin{equation}\label{spa sparity}
    \begin{aligned}        
       {{\rm supp}\{\boldsymbol{\mathscr{E}}_{:,:,1}\}}={\rm supp\{\boldsymbol{\mathscr{E}}_{:,:,2}\}}=\cdots={{\rm supp}\{\boldsymbol{\mathscr{E}}_{:,:,N_r}\}}.
    \end{aligned}
\end{equation}
The set of $\{\boldsymbol{\mathscr{E}}_{:,:,n}\}^{N_r}_{n=1}$ exhibits spatial-frequency structured sparsity, which is jointly defined by (\ref{fre sparity}) and (\ref{spa sparity}). For GF-RA, the pilot phase is used to estimate the AUS $\mathcal{K}_a$ and the associated channel matrices $\{\boldsymbol{\mathscr{E}}_{:,:,n}\}_{n=1}^{N_r}$ from noisy measurements $\{\boldsymbol{\mathscr{Y}}^b_{:,:,n}\}^{N_r}_{n=1}$. We denote the received signal tensor across all antennas by $\boldsymbol{\mathscr{Y}}^b\in\mathbb{C}^{{T}\times G\times N_r}$.

Additionally, the large satellite-terrestrial distance leads to an extremely narrow angular spread, and the link is dominated by the LoS path with very few effective scatterers, resulting in strong sparsity and cluster characteristics in the virtual AD-domain. We define the equivalent virtual AD-domain channel tensor as $\boldsymbol{\mathscr{D}}=\boldsymbol{\mathscr{E}}\!\times\!_3\mathbf{\Theta}^*_{A}\!\times\!_2\mathbf{\Theta}^*_{F}\in\mathbb{C}^{K\times G\times N_r}$, where $\mathbf{\Theta}_A=\mathbf{\Theta}_X\otimes\mathbf{\Theta}_Y\in\mathbb{C}^{N_r\!\times\! N_r}$, $\mathbf{\Theta}_X\in\mathbb{C}^{N^x_r\!\times\! N^x_r}$, $\mathbf{\Theta}_Y\in\mathbb{C}^{N^y_r\!\times\! N^y_r}$, and $\mathbf{\Theta}_F\in\mathbb{C}^{G\!\times\! G}$ are normalized unitary discrete Fourier transform (DFT) matrices. 

The received signal tensor $\boldsymbol{\mathscr{R}}^b\!\in\!\mathbb{C}^{T\times G\times N_r}$ in this domain is
\begin{equation}\label{R_tensor}
    \begin{aligned}         \boldsymbol{\mathscr{R}}^b&=\boldsymbol{\mathscr{Y}}^b\times_3\mathbf{\Theta}^*_{A}\times_2\mathbf{\Theta}^*_F   =\boldsymbol{\mathscr{D}}\times_1\mathbf{X}^b+\widetilde{\boldsymbol{\mathscr{N}}},
    \end{aligned}
\end{equation}
where $\widetilde{\boldsymbol{\mathscr{N}}}=\boldsymbol{\mathscr{N}}\times_3\mathbf{\Theta}^*_{A}\times_2\mathbf{\Theta}^*_F$. Slicing (\ref{R_tensor}) along the angular dimension yields the per-angular form
\begin{equation}\label{R_matrix}
    \begin{aligned}
\boldsymbol{\mathscr{R}}^b_{:,:,n}=\mathbf{X}^b\boldsymbol{\mathscr{D}}_{:,:,n}+\widetilde{\boldsymbol{\mathscr{N}}}_{:,:,n}, n=1,2,\cdots,N_r,
    \end{aligned}
\end{equation}
where $\boldsymbol{\mathscr{R}}^b_{:,:,n}\in\mathbb{C}^{T\times G}$ denotes the received signal slice, $\boldsymbol{\mathscr{D}}_{:,:,n}\in\mathbb{C}^{K\times G}$ is the equivalent channel for the $n$-th virtual angular bin, and $\widetilde{\boldsymbol{\mathscr{N}}}_{:,:,n}$ is the noise slice, which preserves the statistical properties of the original noise. 
\begin{algorithm}[!t]	 
	\SetAlCapFnt{\scriptsize}
	\SetAlCapNameFnt{\scriptsize}
	\begin{scriptsize}
		\captionsetup{font=scriptsize}
		\caption{Proposed IRF-MAMP Algorithm}
		\label{IRF-MMV-AMP}
		\LinesNumbered
		\KwIn{$\forall n$:  $\boldsymbol{\mathscr{Y}}^p_{:,:,n}$, $\mathbf{X}^p$, the maximum number of iterations $L_{\rm iter}$.}
		\KwOut{The estimated AUS $\widehat{\mathcal{K}_a}$ and the channel matrix$\{\widehat{\boldsymbol{\mathscr{E}}}_{:,:,n}\}_{n=1}^{N_r}$.}
		Initialize: $i$=1, $\Lambda^0=\emptyset$, $(\overline{\boldsymbol{\mathscr{Y}}}^p_{:,:,n})^1=\boldsymbol{\mathscr{Y}}^p_{:,:,n}$\;
		\Repeat{$i\! > \!L_{\rm iter}$ {\rm or} $\|(\overline{\boldsymbol{\mathscr{Y}}}^p_{:,:,n})^{i+1}\|^2<10^{-4}$}{		
			$k = 0$, $\Sigma = \Pi = \emptyset$\;
			$\forall n,k,g$: Obtain the posterior belief probability $\eta_{n,k,g}^i$ by applying the MAMP algorithm to model (\ref{Y=XE});	\tcp{$\triangleright$ detection stage}	
			\For{$k=1,2,\cdots, K$}{
				\If {$(GN_r)^{-1}\sum_g{\sum_n\Upsilon\{\eta_{n,k,g}>\epsilon_{\rm low}\}\geq0.9}$} {
					$\Sigma  = \Sigma \cup \Lambda^{i-1} \cup \left\{k\right\}$\;}
				\If {$(GN_r)^{-1}\sum_g{\sum_n\Upsilon\{\eta_{n,k,g}>\epsilon_{\rm high}\}\geq0.9}$}{
					$\Lambda^i  = \Lambda^{i-1}  \cup \left\{k\right\}$
					\;}}	
			$\forall n$: Obtain the channel vector $\widehat{\boldsymbol{\mathscr{D}}}^i_{k,:,n}, \forall k\in \Sigma$ by applying the MAMP algorithm to model (\ref{R_stage})\label{R=XD}; \tcp{$\triangleright$ estimation stage}	
			Acquire set $\Pi$, $\Pi \subseteq \Lambda^i$, and $\left|\Pi\right|_c / \left|\Lambda^i\right|_c = \zeta_{\rm act}$\label{Pi}\;
			$\widehat{\boldsymbol{\mathscr{E}}}^i=\widehat{\boldsymbol{\mathscr{D}}}^i\times_3\mathbf{\Theta}_A\times_2\mathbf{\Theta}_F$, $\left(\overline {\boldsymbol{\mathscr{Y}}}_{:,:,n}^p\right)^{i+1}=\boldsymbol{\mathscr{Y}}^p_{:,:,n}-{\bf X}^p_{:,\Pi} \widehat {\boldsymbol{\mathscr{E}}}_{\Pi,:,n}^i$\label{Ei,Yi+1}\;
			$i = i + 1$\;
		}			
		\KwRet${\widehat {\mathcal K}_a} \gets \Sigma$, $\widehat{\boldsymbol{\mathscr{E}}}_{:,:,n}^{i-1}$.\			
	\end{scriptsize}
\end{algorithm}
\section{Proposed Receiver Detection Scheme at The LEO Satellite}
This section proposes a receiver detection scheme with two core modules: an alternating AUD and CE (Alt-ADCE) module and a DD module. The Alt-ADCE module includes a detection stage for AUD in the SF-domain and an estimation stage for CE in the AD-domain. In the pilot phase, to fully leverage the structured sparsity of $\{\boldsymbol{\mathscr{E}}_{:,:,n}\}_{n=1}^{N_r}$ and the enhanced clustered sparsity of $\{\boldsymbol{\mathscr{D}}_{:,:,n}\}_{n=1}^{N_r}$, we formulate the AUD and CE problems as a compressive sensing problem~\cite{2026_SCIS}, solved by the proposed residual-driven algorithm via alternating iterations between the two stages. Subsequently, in the data transmission phase, we use the estimated channel matrix and received signals at the LEO satellite to perform DD.
\subsection{Residual Feedback-Based Alternating AUD and CE}
To leverage the multi-domain synergistic sparsity gain, we propose the IRF-MAMP algorithm (Alg. \ref{IRF-MMV-AMP}) with detection and estimation stages. When $i=1$, the detection stage processes $\boldsymbol{\mathscr{Y}}^p_{:,:,n}$ using the MAMP algorithm (\ref{MMV-AMP algorithm}) applied to (\ref{spa-fre-Y}) to identify the coarse AUS $\Sigma$ and the high-reliability AUS $\Lambda$ based on posterior active probabilities $\eta_{n,k,g}$, utilizing thresholds $\epsilon_{\rm low}=0.3$ and $\epsilon_{\rm high}=0.9$, respectively. The estimation stage then uses $\Sigma$ to acquire the estimated channel matrix ${\boldsymbol{\mathscr{D}}}_{\Sigma,:,n}$ via MAMP on ($\ref{R_stage}$) as follows
\begin{equation}\label{R_stage}
	\begin{aligned}   \boldsymbol{\mathscr{R}}^p_{:,:,n}=\mathbf{X}^p_{:,\Sigma}{\boldsymbol{\mathscr{D}}}_{\Sigma,:,n}+\mathbf{\overline{\boldsymbol{\mathscr{N}}}}_{:,:,n},
	\end{aligned}
\end{equation}
where $\mathbf{X}^p_{:,\Sigma}\in\mathbb{C}^{{T_p}\!\times\! |\Sigma|_c}$, ${\boldsymbol{\mathscr{D}}}_{\Sigma,:,n}\in\mathbb{C}^{|\Sigma|_c\!\times\! G}$, and $\mathbf{\overline{\boldsymbol{\mathscr{N}}}}_{:,:,n}=\mathbf{X}^p_{:,\mathcal{K}\!-\!\Sigma}{\boldsymbol{\mathscr{D}}}_{\mathcal{K}\!-\!\Sigma,:,n}+\widetilde{\boldsymbol{\mathscr{N}}}_{:,:,n}$. 

The core of this algorithm lies in the feedback of residuals after each iteration. We leverage the estimated reliable UT set $\Lambda$ and the high-precision estimated channel  $\widehat{\boldsymbol{\mathscr{E}}}_{:,:,n}^i$ (transformed back to the SF-domain via $\widehat{\boldsymbol{\mathscr{E}}}^i=\widehat{\boldsymbol{\mathscr{D}}}^i\times_3\mathbf{\Theta}_A\times_2\mathbf{\Theta}_F$) to reconstruct the signal for UTs in $\Pi$. This reconstructed signal is subtracted from the previous total received signal to generate the residual $\left(\overline {\boldsymbol{\mathscr{Y}}}_{:,:,n}^p\!\right)^{\!i+1}\!=\!\boldsymbol{\mathscr{Y}}^p_{:,:,n}-{\bf X}^p_{:,\Pi} \widehat {\boldsymbol{\mathscr{E}}}_{\Pi,:,n}^i$ for the next iteration. To prevent divergence, only signals from a subset $\Pi$ of the reliable set $\Lambda$, determined by the proportion $\zeta_{\rm act}$, are removed.

\begin{figure}[t]
	\centering 
	\captionsetup{font=footnotesize}
	\includegraphics[width=0.9\linewidth]{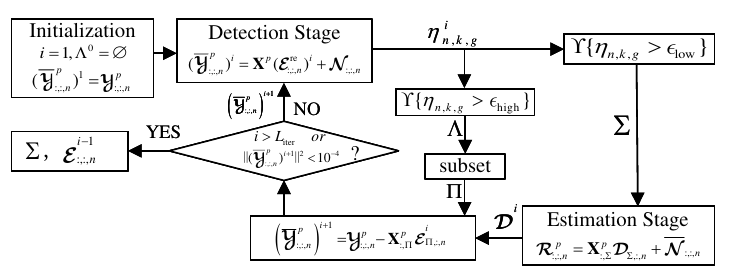}
	\caption{Block diagram of the proposed IRF-MAMP algorithm.}
	\label{framework}
	\vspace{-3pt}
\end{figure}

In subsequent iterations ($i\!>\!1$), the detection stage aims to detect active UTs from the remaining received signals $(\overline{\boldsymbol{\mathscr{Y}}}_{:,:,n}^p)^{i}$ and the estimation stage aims to recover $(\boldsymbol{\mathscr{E}}_{:,:,n}^{\rm re})^i\in\mathbb{C}^{|\mathcal{K}|_c\times G}$. This term is constructed as $(\boldsymbol{\mathscr{E}}_{:,:,n}^{\rm re})^i\!=\!\boldsymbol{\mathscr{E}}_{:,:,n}\!-\!\widetilde{\boldsymbol{\mathscr{E}}}_{:,:,n}^i$, while $\widetilde{\boldsymbol{\mathscr{E}}}_{\Pi,:,n}^i\!=\!\widehat{\boldsymbol{\mathscr{E}}}_{\Pi,:,n}^{i-1}\in\mathbb{C}^{|\Pi|_c\times G}$ and $\widetilde{\boldsymbol{\mathscr{E}}}^i_{\mathcal{K}\!-\!\Pi,:,n}\!=\!\mathbf{0}_{|\mathcal{K}\!-\!\Pi|_c\!\times\! G}$. The detection stage is based on the following model
\begin{equation}\label{Y=XE}
	\begin{aligned}        (\overline{\boldsymbol{\mathscr{Y}}}_{:,:,n}^p)^{i}=\mathbf{X}^p (\boldsymbol{\mathscr{E}}_{:,:,n}^{\rm re})^i + \boldsymbol{\mathscr{N}}_{:,:,n},
	\end{aligned}
\end{equation}
Based on this, the detector can more accurately identify previously missed UTs and update $\Sigma$ as well as $\Lambda$. The updated UT sets are immediately sent to the estimation stage to obtain a more refined channel estimate. The detection and estimation stages are executed alternately until the convergence condition is met as illustrated in Fig.~\ref{framework}.
\subsection{MAMP Algorithm}\label{MMV-AMP algorithm}
To handle the sparse signal recovery problem, we adopt the AMP algorithm. We define $e_{k,g}=\boldsymbol{\mathscr{E}}_{k,g,n}$, and model a spike-and-slab prior distribution for $e_{k,g}$ (omitting the subscript $n$, ${T_p}$, and the superscript $p$ below), i.e.,
\begin{equation}\label{spike-and-slab}
    \begin{aligned}
        p_0(e_{k,g})=(1-\psi_{k,g})\delta(e_{k,g})+\psi_{k,g}\mathcal{CN}(e_{k,g};\mu,\tau),
    \end{aligned}
\end{equation}
where $\psi_{k,g}$ is the sparsity ratio, that is, the probability that $e_{k,g}$ is non-zero.

In this formulation, the minimum mean square error (MMSE) estimate of $\boldsymbol{\mathscr{E}}_{:,:,n}$ coincides with its posterior mean. Although the joint posterior distribution $p(e_{k,g}|\boldsymbol{\mathscr{Y}}_{:,:,n})$ can be computed according to the Bayes' rule, the direct computation is complex. Therefore, we exploit a factor-graph representation of the posterior factorization and perform efficient approximate inference via the AMP algorithm. The update rules for the variance $V^{u}_{t,g}$ and mean $Z^{u}_{t,g}$ at factor nodes in the $u$-th iteration are as follows
\begin{equation}\label{V-Z}
    \begin{split}
        V^{u}_{t,g}&=\!\sum\nolimits_{k}|x_{t,k}|^2v^{u}_{k,g},\\
        Z^{u}_{t,g}&= \sum\nolimits_{k}x_{t,k}\hat{e}^u_{k,g} \!-\! V^{u}_{t,g}(\mathscr{Y}_{t,g}\!-\!Z^{u\!-\!1}_{t,g})/(\sigma^2 + V^{u\!-\!1}_{t,g}),
    \end{split}
 \raisetag{1pt}  
\end{equation}
where $\hat{e}_{k,g}$ denotes the posterior mean of $e_{k,g}$, $v_{k,g}^{u}$ is the posterior variance, $\sigma^2$ is the variance of the AWGN, and $\mathscr{Y}_{t,g}=\boldsymbol{\mathscr{Y}}_{t,g,n}$. The update rules of the variance and mean at variable nodes in the $u$-th iteration are as follows
\begin{equation}\label{D-C}
    \begin{aligned}
        D_{k,g}^{u}&\!=\left[\sum\nolimits_{t}|x_{t,k}|^2/(\sigma^2+V^{u}_{t,g})\right]^{-1},\\
        C^{u}_{k,g}&\!=\!\hat{e}^u_{k,g}\!+\!D_{k,g}^{u}\sum\nolimits_{t}[x^*_{t,k}(\mathscr{Y}_{t,g}\!-\!Z^{u}_{t,g})]/[\sigma^2\!+\!V^{u}_{t,g}].
    \end{aligned}
\end{equation}

The posterior distribution of $e_{k,g}$ is the product of the prior distribution and the message of the variable node, and it can be expressed as 
\vspace{-0.6em}
\begin{equation}\label{scalar}
    \begin{aligned}
        p(e_{k,g}|\boldsymbol{\mathscr{Y}}_{:,:,n})&\approx p(e_{k,g}|C^u_{k,g},D^u_{k,g})\\
        &\approx p_0(e_{k,g})\mathcal{CN}(e_{k,g};C^u_{k,g},D^u_{k,g}).
    \end{aligned}
\end{equation}

Accordingly, the MMSE estimate of the high-dimensional joint posterior distribution $p(e_{k,g}|\boldsymbol{\mathscr{Y}}_{:,:,n})$ is approximately decoupled into a set of scalar posteriors in (\ref{scalar}). This decoupling avoids high-dimensional integrals and thus simplifies the computation of the MMSE estimate. By exploiting the prior model of $p_0(e_{k,g})$ in (\ref{scalar}), the posterior distribution of $e_{k,g}$ is obtained as follows
\begin{equation}\label{posterior}
	\begin{aligned}
       p(e_{k,g}|C^u_{k,g},D^u_{k,g})=&(1-\eta^u_{k,g})\delta(e_{k,g})\\&+\eta^u_{k,g}\mathcal{CN}(e_{k,g};A^u_{k,g},B^u_{k,g}),
	\end{aligned}
\end{equation}
where $		\eta^u_{k,g}=\psi_{k,g}/[\psi_{k,g}+(1-\psi_{k,g}){\rm exp}(-\mathcal{V})]$, and $\eta^u_{k,g}$ is the belief indicator. Following the method in~\cite{2020_JADCE_kml}, we derive $\psi_{k,g}$, $A^u_{k,g}$, $B^u_{k,g}$, and $\mathcal{V}$, as well as initialize and recursively update the unknown hyper-parameters $\{\mu,\tau,\sigma^2,\psi_{k,g}\}$ via the expectation-maximization algorithm. Additionally, we propose a detector using the $\eta_{n,k,g}$, given by $\hat{\alpha}_k=\Upsilon\!\{\frac{1}{GN_r}\sum_{g}\sum_{n}
\Upsilon\!\{\,\eta_{n,k,g}>\epsilon\,\}\ge 0.9\}$, where $\Upsilon\{\cdot\}$ is the indicator function that equals $1$ if the condition holds, and $0$ otherwise.

\begin{figure*}[t]
	\centering 
	\captionsetup{font=footnotesize}
	\includegraphics[width=0.9\linewidth]{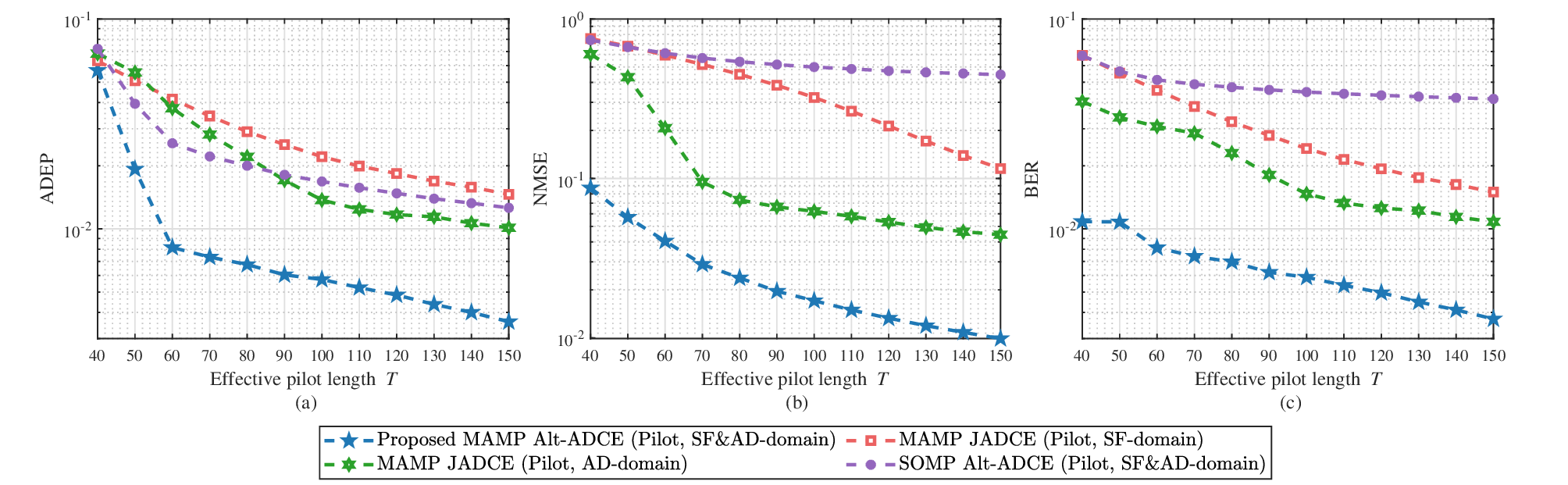}
	\caption{Performance for different schemes versus pilot length at $\mathrm{SNR}=16~\mathrm{dB}$. (a) ADEP. (b) NMSE. (c) BER.}
	\label{difT}
\end{figure*}

\begin{figure}[t]
	\centering 
	\captionsetup{font=footnotesize}
	\includegraphics[width=0.85\linewidth]{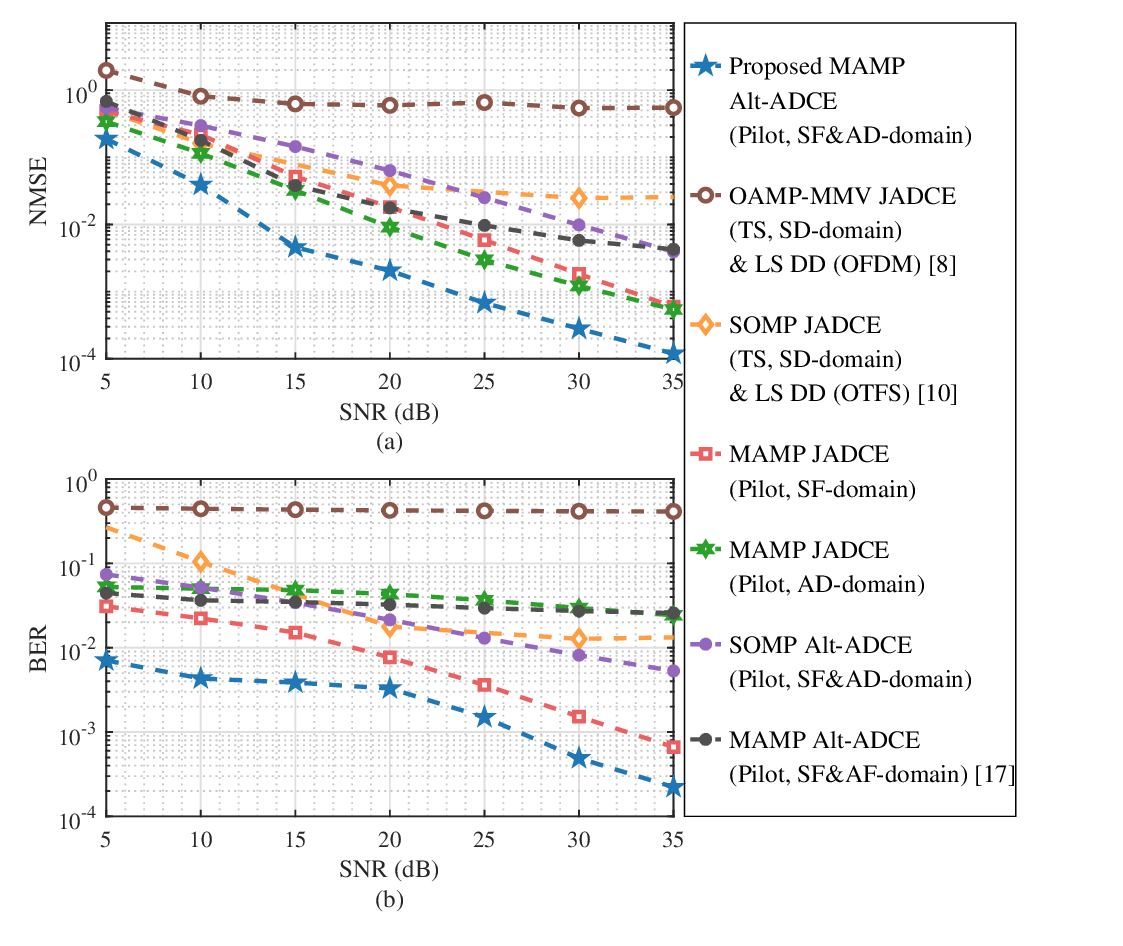}
	\caption{Performance for different schemes versus SNR at $T=80$. (a) NMSE. (b) BER.}
	\label{difSNR}
\end{figure}

\subsection{Data Detection Stage}
After obtaining the estimated channel tensor $\widehat{\boldsymbol{\mathscr{E}}}_{:,:,n}$ from Alt-ADCE, the data symbols are estimated using the received data from all antennas at the LEO satellite. The signal model $\boldsymbol{\mathscr{Y}}^d_{:,:,n}\in\mathbb{C}^{T_d\times G}$ for DD can be expressed as follows
\begin{equation}\label{R}
    \begin{aligned}        \boldsymbol{\mathscr{Y}}^d_{:,:,n}=\mathbf{X}^d\widehat{\boldsymbol{\mathscr{E}}}_{:,:,n}+\widetilde{\boldsymbol{\mathscr{N}}}_{:,:,n},\quad n=1,2,\cdots,N_r,
    \end{aligned}
\end{equation}
where $\mathbf{X}^d\!\in\!\mathbb{C}^{T_d\!\times\! |\widehat{\mathcal{K}_a}|_c}$ denotes the data symbols, and $\widehat{\boldsymbol{\mathscr{E}}}_{:,:,n}\!\in\!\mathbb{C}^{|\widehat{\mathcal{K}_a}|_c\!\times\! G}$ denotes the estimated SF-domain channel matrix of active UTs. By stacking the signals from all satellite antennas, we obtain
\begin{equation}\label{stackR}
    \begin{aligned}
        \mathbf{Y}_{\rm DD}=\mathbf{X}^d\mathbf{\widehat{E}}_{\rm DD}+\widetilde{\mathbf{N}}_{\rm DD},
    \end{aligned}
\end{equation}
where $\mathbf{Y}_{\rm DD}\!=\![\boldsymbol{\mathscr{Y}}^d_{:,:,1},\boldsymbol{\mathscr{Y}}^d_{:,:,2},\cdots,\boldsymbol{\mathscr{Y}}^d_{:,:,N_r}]\!\in\!\mathbb{C}^{T_d\!\times\!GN_r}$, $\mathbf{\widehat{E}}_{\rm DD}=[\widehat{\boldsymbol{\mathscr{E}}}_{:,:,1},\widehat{\boldsymbol{\mathscr{E}}}_{:,:,2},\cdots,\widehat{\boldsymbol{\mathscr{E}}}_{:,:,N_r}]\in\mathbb{C}^{|\widehat{\mathcal{K}_a}|_c\!\times\! GN_r}$, and $\widetilde{\mathbf{N}}_{\rm DD}=[\widetilde{\boldsymbol{\mathscr{N}}}_{:,:,1}, \widetilde{\boldsymbol{\mathscr{N}}}_{:,:,2},\cdots,\widetilde{\boldsymbol{\mathscr{N}}}_{:,:,N_r}]\in\mathbb{C}^{{T_d}\!\times\! GN_r}$. Finally, we estimate $\mathbf{X}^d$ using the LMMSE estimator as follows
\begin{equation}\label{DD}
    \begin{aligned}
        (\widehat{\mathbf{X}}^d)^{\mathsf{T}}=(\mathbf{\widehat{E}}_{\rm DD}\mathbf{\widehat{E}}_{\rm DD}^{\mathsf{H}}+\hat{\sigma}^2\mathbf{I}_{|\widehat{\mathcal{K}_a}|_c})^{-1}\mathbf{\widehat{E}}_{\rm DD}\mathbf{Y}^{\mathsf{T}}_{\rm DD}.
    \end{aligned}
\end{equation}
\subsection{Computational Complexity Analysis}
In view of the requirements for hardware cost and power consumption in systems with massive access, this section analyzes the computational complexity of the proposed algorithm, covering the detection stage, estimation stage, and DD stage.
	
In each outer residual-feedback iteration, the MAMP algorithm operates on all $K$ potential users in the detection stage, and its computational complexity over $L_{\rm amp}$ inner MAMP iterations is $\mathcal{O}(L_{\rm amp} \times (4 T_p K G N_r + 3 T_p K G + 16 T_p G N_r + 20 K G N_r))$. In the estimation stage, the algorithm is performed only on the coarse AUS $\Sigma$, so the computational complexity per iteration is $\mathcal{O}(L_{\rm amp} \times (4 T_p |\Sigma|_c G N_r + 3 T_p |\Sigma|_c G + 16 T_p G N_r + 20 |\Sigma|_c G N_r))$, where $L_{\rm amp}$ is the maximum number of iterations. Therefore, the overall Alt-ADCE complexity over $L_{\rm iter}$ outer residual-feedback iterations is obtained by multiplying the above detection and estimation complexities by $L_{\rm iter}$. The computational complexity of the DD stage is $\mathcal{O}(|\hat{\mathcal{K}}_a|_c^3 + 2|\hat{\mathcal{K}}_a|_c^2 G N_r + |\hat{\mathcal{K}}_a|_c G N_r T_d)$. For comparison, the OAMP-MMV-based scheme in~\cite{2023_Ying_LEO} has a dominant JADCE complexity of approximately $\mathcal{O}(T_p^2 K G L_{\rm oamp}+3T_p K G N_r L_{\rm oamp})$, where the $L_{\rm oamp}$ is the maximum number of iterations, followed by channel refinement depending on the active UT number and array-processing dimension. The SOMP-based scheme in~\cite{2023_zxy_LEO} is mainly dominated by greedy correlation search and support-set updating, with an approximate upper bound of $\mathcal{O}(2T_p (N_{\rm OTFS}+1) N_r KG+KG(N_{\rm OTFS}+1)N_r+2L_{\rm somp}^2 T_p+L_{\rm somp}^3+L_{\rm somp}T_p(N_{\rm OTFS}+1)N_r)$, where $N_{\rm OTFS}$ and $L_{\rm somp}$ denote the Doppler-domain dimension of the OTFS frame and the number of SOMP iterations, respectively. Compared with these methods, the proposed IRF-MAMP requires moderate additional receiver-side computation and memory due to residual feedback, but improves robustness in overloaded massive access scenarios, as verified in Section~\ref{Performance Evaluation}.

\section{Performance Evaluation}
\label{Performance Evaluation}
We verify the effectiveness of the proposed scheme through simulations. We consider a typical massive access scenario with $K=500$ potential UTs and $K_a=50$. The LEO satellite operates at an altitude of $550$ km. Other simulation parameters are as follows: carrier frequency $f_c=14.5$ GHz, subcarrier spacing $480$ kHz, $512$ subcarriers, and system bandwidth $245.76$ MHz. The number of channel paths is set to $L_p=1$, and the cyclic prefix length is $36$ samples. The maximum support of the channel impulse response consists of $8$ taps, spanning $0.02848$ $\rm \mu s$. The power distribution factor is $\gamma=10$ dB, and the pulse shaping filter is $\varrho(\tau)=\delta(\tau)$. Unless otherwise specified, each SG has $G=16$ subcarriers and $N_r^x=N_r^y=5$. The spreading code $\mathbf{S}$ is constructed using deterministic DFT-based constant-modulus complex exponential sequences.

For performance evaluation, we use the activity detection error probability (ADEP), the normalized mean square error (NMSE) between the true channel matrix and the estimated channel matrix, and the bit error rate (BER) as the metrics of AUD, CE, and DD, respectively. We adopt six GF-RA baselines for comparison with our \textit{Proposed MAMP Alt-ADCE (Pilot, SF$\&$AD-domain)} (hereafter denoted as the \textbf{Proposed Scheme}). \textit{OAMP-MMV JADCE (TS, SD-domain) $\&$ LS DD (OFDM)} (denoted as \textbf{Baseline 1}): This baseline utilizes the training sequence (TS) padded frame structure from~\cite{2023_Ying_LEO}, performing JADCE via OAMP-MMV in the spatial-delay domain (SD-domain) and using the least squares (LS) method for DD. \textit{SOMP JADCE (TS, SD-domain) $\&$ LS DD (OTFS)} (denoted as \textbf{Baseline 2}): This scheme~\cite{2023_zxy_LEO} uses the TS and the SOMP algorithm for JADCE in the SD-domain, and LS method for DD (to ensure fairness, the non-inter-symbol-interference region in~\cite{2023_Ying_LEO} and~\cite{2023_zxy_LEO} is set to the effective pilot length). \textit{MAMP JADCE (Pilot, SF-domain)} (denoted as \textbf{Baseline 3}): This baseline uses a joint spatial-frequency spreading data modulation scheme, performing JADCE with MAMP in the SF-domain and LMMSE for DD. \textit{MAMP JADCE (Pilot, AD-domain)} (denoted as \textbf{Baseline 4}): This baseline is similar to Baseline 3, but performs JADCE in the AD-domain. \textit{SOMP Alt-ADCE (Pilot, SF$\&$AD-domain)} (denoted as \textbf{Baseline 5}): This baseline differs from the Proposed scheme by using the SOMP algorithm instead of AMP. \textit{MAMP Alt-ADCE (Pilot, SF$\&$AF-domain)} (denoted as \textbf{Baseline 6}): This baseline~\cite{2020_JADCE_kml} differs from the Proposed Scheme by omitting the joint spatial-frequency spreading and performing alternating iterations between the spatial-frequency and angular-frequency (AF) domains.

Since the compared schemes employ different frame structures, we define  \textit{Effective pliot length} as the number of interference-free temporal observations used for sparse recovery. For the Proposed Scheme and Baselines 3–6, where each temporal observation corresponds to one pilot OFDM time slot. For the TS-padded Baselines 1 and 2, denotes the length of the non-inter-symbol-interference region extracted from the received TS in time-domain samples, rather than the total TS length. Therefore, the comparisons equalize the effective measurement dimension of the JADCE problem, but do not imply identical waveform durations or physical pilot overheads across different frame structures.

Fig.~\ref{difT} illustrates the ADEP, NMSE, and BER performance comparisons for the different schemes under varying effective pilot lengths $T$. It is evident that the Proposed scheme demonstrates significant performance superiority across various evaluation metrics. Specifically, as illustrated in Fig.~\ref{difT}(b) and Fig.~\ref{difT}(c), Baseline 4 outperforms Baseline 3 by leveraging the clustered sparsity inherent in the angular-delay domain, whereas the latter relies solely on SF domain sparsity. Furthermore, due to the inability to fully exploit prior information, Baseline 5 exhibits inferior performance, and its NMSE and BER show negligible improvement within the illustrated range of effective pilot lengths. Notably, the Proposed scheme maintains robust access performance even under low pilot overhead regimes.

Fig.~\ref{difSNR} shows the performance of different schemes versus signal-to-noise ratio (SNR) at $T=80$, where Fig.~\ref{difSNR}(a) and Fig.~\ref{difSNR}(b) present the NMSE and BER results, respectively. As the SNR increases, both the NMSE and BER of all schemes exhibit the expected downward trend. The \textit{Proposed Scheme} consistently outperforms all baselines over the entire plotted SNR range, achieving the lowest estimation error in Fig.~\ref{difSNR}(a) and the lowest BER in Fig.~\ref{difSNR}(b), which demonstrates its clear advantages in both channel estimation and data detection. In contrast, Baseline 3 and Baseline 4 perform JADCE only in a single domain and therefore suffer from the lack of multi-domain synergistic gain. Although Baseline 6 also adopts alternating processing over two domains, it still performs worse than the \textit{Proposed Scheme}, since it does not employ the proposed joint spatial-frequency spreading design and thus cannot alleviate the underdetermined demodulation problem by expanding the effective observation dimension. It is worth noting that, due to the severe path loss commonly encountered in satellite-to-ground links, practical LEO satellite communication systems often operate in the low-to-medium SNR region, making this operating regime of greater practical relevance.

\begin{figure}[t]
	\centering 
	\captionsetup{font=footnotesize}
	\includegraphics[width=0.85\linewidth]{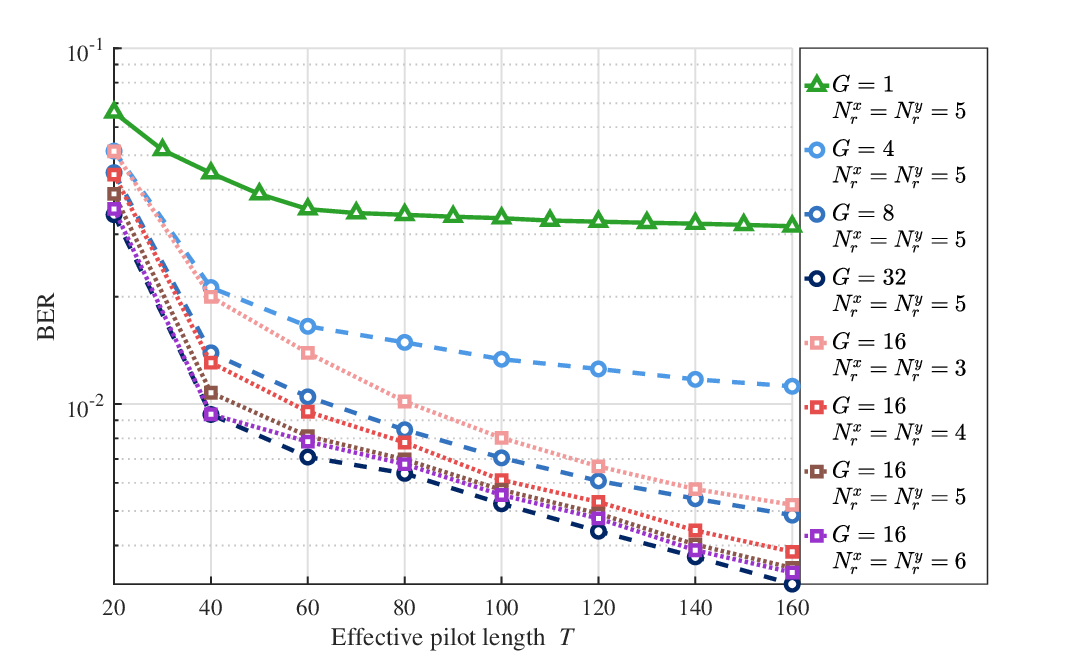}
	\caption{DD performance of the proposed scheme for different numbers of subcarriers and satellite antennas at $\mathrm{SNR}=16~\mathrm{dB}$.}
	\label{difAnteSubCar}
\end{figure}

Fig.~\ref{difAnteSubCar} illustrates the trend of BER performance versus effective pilot length $T$ at $\mathrm{SNR}=16~\mathrm{dB}$. First, the Proposed scheme that utilizes the joint spatial-frequency spreading data modulation scheme ($G>1$) significantly outperforms the baseline non-spreading scheme ($G=1$), thereby validating the effectiveness of expanding the signal observation dimension to overcome the physical limitation of the number of satellite antennas. Regarding the scenario with a fixed antenna scale of $N_r=5\times5$, the BER exhibits a continuous downward trend as the number of subcarriers $G$ increases. Specifically, the brown curve with $G=16$ outperforms the $G=8$ case, while the configuration of $G=32$ retains the best performance due to fuller exploitation of frequency-domain sparsity gains. Similarly, for the schemes with a fixed number of subcarrier $G=16$, increasing the number of satellite antennas $N_r$ leads to a monotonic decrease in BER, as the system benefits from enhanced array gain and spatial diversity. However, it is observed that the performance gap between the $N_r=5\times5$ and $N_r=6\times6$ cases gradually narrows at longer pilot lengths, indicating that the performance gain tends to saturate, while the $N_r=6\times6$ configuration still achieves superior performance.

\begin{figure}[t]
	\centering 
	\captionsetup{font=footnotesize}
	\includegraphics[width=0.9\linewidth]{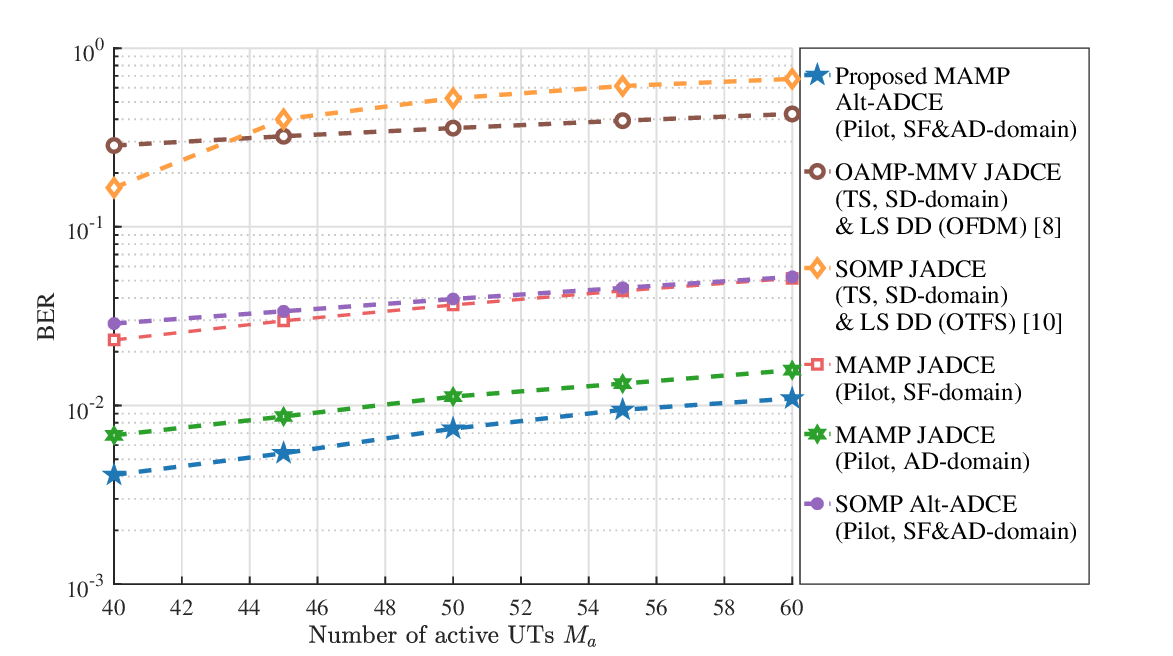}
	\caption{DD performance of different schemes with varying numbers of active UTs, with $T=60$ and $\mathrm{SNR}=20~\mathrm{dB}$.}
	\label{difActUe}
\end{figure}

Fig.~\ref{difActUe} illustrates the DD performance of different schemes versus the number of active UTs $K_a$ given $T=60$ and $\mathrm{SNR}=20~\mathrm{dB}$. As $K_a$ increases from 40 to 60, the BERs of all schemes increase, with Baseline 2 and Baseline 1 exhibiting the worst performance and fastest degradation. Baseline 1 relies only on common antenna support, which leads to a highly correlated multi-user channel matrix, while its zero-forcing detection further amplifies the estimation error. Baseline 2 suffers from both insufficient spatial sparsity utilization and performance degradation as SOMP is highly sensitive to the measurement matrix column correlation caused by proximate AoAs as $K_a$ increases. Furthermore, Baseline 3 and Baseline 5 perform similarly, and both are inferior to Baseline 4, which indicates that methods lacking multi-domain synergistic gain or relying on greedy algorithms fail to effectively suppress error accumulation under high UT loads. 

\section{Conclusions} 
This paper investigates the challenge of massive connectivity in LEO satellite networks and proposes a comprehensive detection framework. Specifically, the developed IRF-MAMP algorithm adopts a multi-domain synergistic strategy that alternately exploits structured sparsity in the SF-domain for AUD and cluster sparsity in the AD-domain for CE, while a residual feedback mechanism is introduced to effectively suppress error accumulation. Furthermore, to address the rank-deficient demodulation issue caused by the underdetermined system, a data modulation scheme with joint SF-domain spreading is designed to effectively expand the signal observation dimension. It is worth noting that, although the proposed spatial-frequency spreading improves system observability and BER performance, it also introduces additional synchronization, computational, and memory costs at the receiver. Therefore, the proposed scheme should be viewed as a trade-off between performance gains and implementation costs, and a moderate spreading dimension is generally more practical under realistic satellite payload constraints.

\bibliographystyle{IEEEtran}
\bibliography{sample}

\end{document}